\documentclass[prl,reprint,superscriptaddress,preprintnumbers]{revtex4-1}

\newif\ifpublic\publictrue

\usepackage{fancyhdr}
\ifpublic\else
\fi
\newif\ifworking\workingtrue

\usepackage{mathtools,mathrsfs,amsbsy,amssymb,latexsym,amsfonts,amscd,
	amsmath, multirow}
\usepackage{bm}

\usepackage{graphicx}
\usepackage[usenames,dvipsnames]{color}
\usepackage[normalem]{ulem}
\usepackage{natbib}
\usepackage{xcolor}
\usepackage{comment}
\definecolor{linkcolor}{rgb}{0,0,0.6}
\usepackage[ pdftex,colorlinks=true,
pdfstartview=FitV,
linkcolor= linkcolor,
citecolor= linkcolor,
urlcolor= linkcolor,
hyperindex=true,
hyperfigures=false]
{hyperref}

\begin{document}
\title{Euclidean flows, solitons and wormholes in AdS from M-theory}

\author{Andr\'es Anabal\'on}
\email{andres.anabalon@uai.cl}
\affiliation{Departamento de Física, Universidad de Concepción,
Casilla 160-C, Concepción, Chile.}
\affiliation{Instituto de F\'isica Te\'orica, UNESP-Universidade Estadual Paulista, R. Dr. Bento T. Ferraz 271, Bl. II, Sao Paulo 01140-070, SP, Brazil}

\author{\'Alvaro Arboleya}
\email{alvaroarbo@gmail.com}
\affiliation{Departamento de F\'isica, Universidad de Oviedo, Avda. Federico Garc\'ia Lorca 18, 33007 Oviedo, Spain}
\affiliation{Instituto Universitario de Ciencias y Tecnolog\'ias Espaciales de Asturias (ICTEA), Calle de la Independencia 13, 33004 Oviedo, Spain}

\author{Adolfo Guarino}
\email{adolfo.guarino@uniovi.es}
\affiliation{Departamento de F\'isica, Universidad de Oviedo, Avda. Federico Garc\'ia Lorca 18, 33007 Oviedo, Spain}
\affiliation{Instituto Universitario de Ciencias y Tecnolog\'ias Espaciales de Asturias (ICTEA), Calle de la Independencia 13, 33004 Oviedo, Spain}

\begin{abstract}

Multi-parametric families of non-supersymmetric EAdS$_{4}$ flows as well as asymptotically EAdS$_{4}$ solitons and wormholes are constructed within the four-dimensional $\textrm{SO}(8)$ gauged supergravity that describes the compactification of M-theory on $\textrm{S}^{7}$. More concretely, the solutions are found within the so-called STU-model that describes the $\textrm{U}(1)^4$ invariant sector of the theory. The on-shell action and gravitational free energy are computed for the regular solutions, the latter being zero for the wormholes irrespective of supersymmetry. There are special loci in parameter space yielding solutions with enhanced (super)symmetry. Examples include a supersymmetric EAdS$_{4}$ flow with $\textrm{SO}(4) \times \textrm{SO}(4)$ symmetry dual to a specific real mass deformation of ABJM on S$^{3}$ as well as a non-supersymmetric wormhole with $\textrm{SU}(3) \times \textrm{U}(1)^2$ symmetry. Uplift formulae for these and other examples to Euclidean solutions of eleven-dimensional supergravity are presented and their complex nature discussed.

\end{abstract}


\maketitle

\section{Introduction}

General Relativity in Euclidean signature has played a central role in understanding thermodynamical aspects of gravitational systems, the most prominent example being the black hole thermodynamics \cite{Gibbons:1976ue}. In the more modern context of the gauge/gravity correspondence \cite{Maldacena:1997re}, Euclidean solutions of supergravity theories have been utilised to explore thermodynamical aspects of strongly-coupled field theories in the planar limit. An example is the holographic evaluation of (the finite part of) the free energy of a three-dimensional conformal field theory (CFT) placed on S$^3$, which has been proposed as a measure of the number of its degrees of freedom. In analogy with the $c$-theorem \cite{Zamolodchikov:1986gt}, this free energy is conjectured to obey a monotonicity property under the renormalisation group (RG) flow, namely, it obeys an $F$-theorem \cite{Jafferis:2011zi,Klebanov:2011gs}.

This note is framed within the specific context of the AdS$_{4}$/CFT$_{3}$ correspondence. On the gravity side, our playground is the four-dimensional (Euclideanised) ${\mathcal{N}=2}$ STU-model of \cite{Cvetic:1999xp} which describes the $\textrm{U}(1)^{4}$ invariant sector of the maximal $\textrm{SO}(8)$ gauged supergravity \cite{deWit:1982ig}. The latter describes the consistent truncation of eleven-dimensional (11D) supergravity on S$^{7}$ to its zero mass sector \cite{deWit:1986oxb}. On the field theory side, the CFT$_{3}$ of relevance is the ABJM theory describing the worldvolume dynamics of a stack of $N$ planar M2-branes in flat space \cite{Aharony:2008ug}. Placing the theory on S$^3$ and turning on $\mathcal{N}=2$ real mass parameters, which modify the assignment of $\textrm{U}(1)_{\textrm{R}}$ R-charges and break conformality, turns out to induce RG flows. Within the STU-model, such RG flows were holographically constructed in \cite{Freedman:2013oja} as a three-parameter family of $\mathcal{N}=2$ Euclidean solutions that preserve the isometries of the S$^3$ in the bulk and have an S$^3$ boundary. Building upon these results, we will construct various new classes of Euclidean solutions within the STU-model: non-backreacted EAdS$_{4}$ flows with non-trivial scalar fields in the bulk, as well as backreacted solutions describing singular solitons and regular wormholes.

Examples of Euclidean four-dimensional solutions with a non-backreacted EAdS$_{4}$ geometry and non-trivial matter fields have previously been constructed in \cite{Nakayama:2016ydc} and uplifted to M-theory in \cite{Nakayama:2016xzs}. These examples involve non-trivial vector fields in the bulk and describe a non-trivial end point of the RG flow induced by topologically twisted (scalar) deformations which preserve the scale invariance but not the conformal invariance of the CFT$_{3}$ in the UV. Here we will present two different three-parameter families -- one singular and one regular -- of non-backreacted EAdS$_{4}$ flows with running scalars (instead of vectors) which are generically non-supersymmetric. For a specific choice of parameters, both the singular and the regular flows become supersymmetric and we present their uplift to 11D supergravity. The supersymmetric and regular EAdS$_{4}$ flow lies within the class constructed in \cite{Freedman:2013oja}. We will review its holographic realisation as a specific real mass deformation of ABJM on S$^{3}$.

In addition to the non-backreacted flows, we also present multi-parametric families of soliton and wormhole solutions of the (Euclideanised) STU-model with running scalars in the bulk. While the solitons are always singular, fully regular wormholes are shown to exist upon judicious choice of the free parameters. The construction of regular Euclidean wormholes in string/M-theoretic scenarios has proven a challenging task. Some constructions for flat-space wormholes have been put forward in the context of (super)gravity coupled to proper-scalars and pseudo-scalars (a.k.a. axions) \cite{Giddings:1987cg,Bergshoeff:2004fq,Bergshoeff:2004pg,Arkani-Hamed:2007cpn}, as well as for AdS wormholes in presence of a cosmological constant or a scalar potential \cite{Gutperle:2002km,Hertog:2017owm,Marolf:2021kjc,Loges:2023ypl,Astesiano:2023iql}. In all these constructions, the analytic continuation from Lorentzian to Euclidean signature becomes subtle as far as the axions are concerned. Fortunately, a full-fledged construction of the Euclidean STU-model was done in \cite{Freedman:2013oja} and shown to require, amongst other modifications, a doubling of the scalar degrees of freedom: the three complex scalars $z_{i}$ of the STU-model and their would-be conjugates $\tilde{z}_{i}$ should be treated as \textrm{independent} fields. This additional freedom in the Euclidean theory is precisely what allows us to construct regular wormholes for which $\tilde{z}_{i} \neq z^{*}_{i}$. When uplifted to eleven-dimensional supergravity, both the 11D metric and the four-form flux become complex-valued with non-trivial real and imaginary parts. Although this is in principle allowed in a classical theory of supergravity, it remains to be seen what the ultimate fate of these Euclidean wormholes will be in a path integral formulation of quantum gravity.

\section{Euclidean supergravity}

The STU-model of \cite{Cvetic:1999xp} describes the $\textrm{U}(1)^4$-invariant sector of the $\textrm{SO}(8)$-gauged maximal supergravity \cite{deWit:1982ig}. Its Euclidean version, which is the relevant one for this work, has been discussed in full detail in \cite{Freedman:2013oja}. In the absence of vector fields, this supergravity describes an Einstein-scalar model involving a set of complex scalars $z^{i}$ and $\tilde{z}^{i}$ with $i=1,2,3$. The bulk Euclidean (bosonic) action is given by
\begin{equation}
\label{S_STU-model}
S_{\textrm{bulk}} = \frac{1}{2\kappa^{2}} \int d^4x \, e \left[ - \frac{R}{2} + \sum_{i=1}^{3} \frac{\partial_{\rho}z_{i}\partial^{\rho}\tilde{z}_{i}}{\left(1-z_{i}\tilde{z}_{i}\right)^{2}}  + V\right] \ ,
\end{equation}
with $2\kappa^2 = 8 \pi G_{4}$ and $G_{4}$ being the four-dimensional Newton constant. The bulk action (\ref{S_STU-model}) includes a non-trivial scalar potential
\begin{equation}
\label{V_STU-model}
V(z_{i},\tilde{z}_{i}) =  g^{2}  \left( 3 - \sum_{i} \frac{2}{1-z_{i}\tilde{z}_{i}}   \right) \ ,
\end{equation}
where $g$ is the four-dimensional gauge coupling that is related to the (inverse) radius of the seven-sphere. Importantly, as emphasised in \cite{Freedman:2013oja}, the scalars $z_{i}$ and $\tilde{z}_{i}$ are complex and independent in the Euclidean theory, and must separately parameterise the Poincar\'e unit-disk, namely, $|z_{i}|, |\tilde{z}_{i}|<1$ $\forall i$. The spacetime metric should also be allowed to be complex in the Euclidean theory, although we are only considering real metrics in this work.

The bulk action,  as specified by (\ref{S_STU-model}) and (\ref{V_STU-model}), possesses three global symmetries $\mathbb{C}_{i}$ acting as constant scalings of the form $z_{i} \rightarrow \lambda_{i} \, z_{i}$ and $\tilde{z}_{i} \rightarrow \lambda^{-1}_{i} \, \tilde{z}_{i}$ with ${\lambda_{i}\in \mathbb{C}}$. However, different four-dimensional solutions related by these symmetries give rise to physically inequivalent backgrounds when uplifted to eleven dimensions (see the uplift formulae collected in Appendix~\ref{app:appendix_11D} and also \cite{Pope:2003jp,Anabalon:2022fti} for a discussion of this issue in the Lorentzian realm).  A proper understanding of global symmetries in the bulk would also require the inclusion of the vector fields in the STU-model as they couple to the scalars through kinetic and topological terms. The $\mathbb{C}_{i}$ symmetries would act linearly on the vector fields changing their boundary values and, therefore, also changing the dual field theory on the boundary. However, we are not considering vector fields in this work and set them to zero.

We will construct non-conformal flow solutions in the Euclidean STU-model that involve non-constant scalars. Moreover, we will demand that the spacetime geometry has an S$^3$ boundary and that the flows preserve the $\textrm{SO}(4) \sim \textrm{SU}(2)_{l} \times \textrm{SU}(2)_{r}$ isometry group of S$^3$ in the bulk. The Euclidean metric (in the Fefferman-Graham gauge) is then of the form
\begin{equation}
\label{ds4_ansatz}
ds_{4}^{2} = d\mu^2 + g^{-2} e^{2A(\mu)}   d\Sigma _{\textrm{S}^{3}}^{2}  \  ,
\end{equation}
where $\mu \in \mathbb{R}$ is the radial (holographic) coordinate and $d\Sigma _{\textrm{S}^{3}}^{2}$ is the line element of a three-sphere $\textrm{S}^{3}$. As a result, the geometry is specified by the metric function $A(\mu)$ which we take to be real-valued. Moreover, in order to preserve the S$^3$ isometries, the scalars can only depend on the radial coordinate, namely, $z_{i}=z_{i}(\mu)$ and $\tilde{z}_{i}=\tilde{z}_{i}(\mu)$. With this ansatz, the second-order equations of
 motion that follow from the bulk action (\ref{S_STU-model}) are given by
\begin{equation}
\label{EOM_4D}
\begin{array}{rcll}
A'' + g^2 e^{-2A} + \displaystyle\sum_{i=1}^{3} \dfrac{z_{i}' \, \tilde{z}_{i}'}{\left( 1- z_{i} \tilde{z}_{i}\right)^2}  & = & 0 & , \\[4mm]
z_{i}'' + 3 A' z_{i}' + 2 \dfrac{\tilde{z}_{i} (z_{i}')^2}{1- z_{i} \tilde{z}_{i}} + 2 g^2 z_{i} & = & 0 & , \\[4mm]
\tilde{z}_{i}'' + 3 A' \tilde{z}_{i}' + 2 \dfrac{z_{i} (\tilde{z}_{i}')^2}{1- z_{i} \tilde{z}_{i}} + 2 g^2 \tilde{z}_{i} & = & 0 & ,
\end{array}
\end{equation}
and are obviously invariant under the $\mathbb{C}_{i}$ scaling symmetries of the bulk action.

The Euclidean STU-model admits an $\mathcal{N}=1$ rewriting in terms of a K\"ahler potential 
\begin{equation}
\label{Kahler_func}
\mathcal{K} = - \sum_{i=1}^{3} \log \left[ 1-z_{i}\tilde{z}_{i} \right] \ ,
\end{equation}
and holomorphic superpotentials
\begin{equation}
\label{W_funcs}
W(z_{i})= g \, (1+z_{1} z_{2} z_{3})
\hspace{3mm} , \hspace{3mm}
\tilde{W}(\tilde{z}_{i}) = g  \, (1+\tilde{z}_{1} \tilde{z}_{2} \tilde{z}_{3}) \ .
\end{equation}
In this language, the bulk action (\ref{S_STU-model}) is constructed from $\mathcal{K}$, $W$ and $\tilde{W}$ using standard $\mathcal{N}=1$ formulae adapted to Euclidean signature (see \cite{Freedman:2013oja,Bobev:2018wbt}), and a set of first-order BPS equations can be derived by requiring the vanishing of the supersymmetry variations of the fermions in the model. The BPS equations can be written as first-order flow equations for the scalars $(z_{i},\tilde{z}_{i})$ and the metric function $A$, with the $\mathcal{N}=1$ gravitino mass $m_{3/2} \propto e^{\frac{\mathcal{K}}{2}} (W \tilde{W})^{\frac{1}{2}}$ playing the role of a scalar superpotential. Using $\mathcal{K}$, $W$ and $\tilde{W}$ in (\ref{Kahler_func}) and (\ref{W_funcs}), the set of BPS equations reads
\begin{equation}
\label{BPS_4D}
\begin{array}{rcll}
-1 + \dfrac{e^{2A}}{g^2} (A')^{2} &=& e^{2A} \dfrac{\left(1 + z_{1}z_{2}z_{3}\right) \left(1 + \tilde{z}_{1}\tilde{z}_{2}\tilde{z}_{3}\right)}{\prod_{i=1}^{3}(1-z_{i} \tilde{z}_{i})} & , \\[5mm]
\dfrac{e^{A}}{g} \dfrac{\left(1 + \tilde{z}_{1}\tilde{z}_{2}\tilde{z}_{3}\right) z_{i}' }{1-z_{i} \tilde{z}_{i}} & = & \left(  \pm 1 -\dfrac{e^{A}}{g} A'\right)  \left( z_{i}  + \dfrac{\tilde{z}_{1}\tilde{z}_{2}\tilde{z}_{3}}{\tilde{z}_{i}} \right) & , \\[4mm]
\dfrac{e^{A}}{g} \dfrac{\left(1 + z_{1}z_{2}z_{3}\right) \tilde{z}_{i}' }{1-z_{i} \tilde{z}_{i}} & = & \left( \mp  1 -\dfrac{e^{A}}{g} A'\right)  \left( \tilde{z}_{i}  + \dfrac{z_{1}z_{2}z_{3}}{z_{i}} \right) & , 
\end{array}
\end{equation}
where the upper (lower) choices of sign correspond to using Killing spinors that are proportional to the left-invariant (right-invariant) Killing spinors on the $\textrm{S}^{3}$ \cite{Freedman:2013oja}. Note that both sign choices are related by the exchange $z_{i} \leftrightarrow \tilde{z}_{i}$. The BPS equations (\ref{BPS_4D}) are generically not invariant under the $\mathbb{C}_{i}$ global scaling symmetries of the bulk action unless two out of the three scalars $z_{i}$ (equivalently for $\tilde{z}_{i}$) are set to zero. While the general flow solutions we will present in this note are non-supersymmetric, and therefore satisfy (\ref{EOM_4D}) without satisfying (\ref{BPS_4D}), we will identify a particular case for which the BPS equations (\ref{BPS_4D}) are additionally satisfied. This supersymmetric case precisely turns off two out of the three complex scalars, \textit{i.e.} $z_{2}=z_{3}=\tilde{z}_{2}=\tilde{z}_{3}=0$, so that the remaining $\mathbb{C}_{1}$ scaling symmetry is also a symmetry of the BPS equations in the bulk.

All the flow solutions we will construct asymptote to the maximally (super) symmetric EAdS$_{4}$ vacuum of the $\textrm{SO}(8)$ supergravity. This vacuum solution uplifts to the $\textrm{EAdS}_{4} \times \textrm{S}^7$ Freund--Rubin background of (Euclidean) eleven-dimensional supergravity with a round metric on the $\textrm{S}^7$  \cite{Freund:1980xh}, and is dual to the superconformal ABJM theory \cite{Aharony:2008ug} placed on S$^3$. This vacuum sits at the origin of field space, namely, ${z_{i}=\tilde{z}_{i}=0}$, and all our solutions will reach it at $|\mu| \rightarrow \infty$ with a fall-off for the scalars of the form
\begin{equation}
\label{scalars_expasions_ab}
\begin{array}{rcll}
z_{i} & = & a_{i} \, e^{-g|\mu|} + b_{i} \, e^{-2g|\mu|} + \ldots & , \\[2mm]
\tilde{z}_{i} & = & \tilde{a}_{i} \, e^{-g|\mu|} + \tilde{b}_{i} \, e^{-2g|\mu|} + \ldots & .
\end{array}
\end{equation}
According to the AdS/CFT correspondence, the fall-off coefficients $(a_{i},b_{i})$ and $(\tilde{a}_{i},\tilde{b}_{i})$ in the expansions (\ref{scalars_expasions_ab}) are related to sources and VEV's in the field theory dual to the supergravity solution. However, the precise identification turns out to be very subtle as the combination $z_{i}-\tilde{z}_{i}$ should be quantised using regular boundary conditions whereas alternate boundary conditions should be used for $z_{i}+\tilde{z}_{i}$. A careful analysis of the boundary limit of the bulk supersymmetry performed in \cite{Freedman:2013oja} yielded the following dictionary. Sources in the dual field theory are identified with the combinations
\begin{equation}
\label{sources_general}
a_{i} - \tilde{a}_{i} 
\hspace{3mm} \textrm{ and } \hspace{3mm}
\left( b_{i} - \frac{\tilde{a}_{1}\tilde{a}_{2}\tilde{a}_{3}}{\tilde{a}_{i}} \right) + \left( \tilde{b}_{i} -\frac{a_{1}a_{2}a_{3}}{a_{i}} \right) \ ,
\end{equation}
whereas VEV's are given by
\begin{equation}
\label{vev_general}
a_{i} + \tilde{a}_{i} 
\hspace{3mm} \textrm{ and } \hspace{3mm}
\left( b_{i} - \frac{\tilde{a}_{1}\tilde{a}_{2}\tilde{a}_{3}}{\tilde{a}_{i}} \right) - \left( \tilde{b}_{i} - \frac{a_{1}a_{2}a_{3}}{a_{i}} \right) \ .
\end{equation}
This identification of conjugate variables (sources and VEV's) is achieved by selecting a holographic renormalisation scheme compatible with supersymmetry. As stated in \cite{Freedman:2013oja}, the counterterms of holographic renormalisation have a universal structure and must be valid for all solutions of the classical field equations of a given bulk theory. We will then follow the holographic renormalisation prescription in \cite{Freedman:2013oja} to compute the on-shell action and gravitational free energy of the supergravity solutions constructed in this work.

\section{Euclidean solutions}

The simplest solution to the system of second order equations in  (\ref{EOM_4D}) is given by a pure EAdS$_{4}$ vacuum of the form
\begin{equation}
\label{EAdS_solution}
\textrm{EAdS$_{4}\,$ :} \hspace{3mm}  e^{2A} = \sinh^2(g\mu)
\hspace{4mm} \textrm{ and } \hspace{4mm} 
z_{i}=\tilde{z}_{i} = 0 \ .
\end{equation}
This solution is maximally supersymmetric within the full $\mathcal{N}=8$ supergravity of \cite{deWit:1982ig}, so it satisfies the BPS equations in (\ref{BPS_4D}). As mentioned before, it uplifts to the $\textrm{EAdS}_{4} \times \textrm{S}^7$ vacuum solution of (Euclidean) eleven-dimensional supergravity with a round metric on the $\textrm{S}^7$ \cite{Freund:1980xh}, and is dual to the superconformal ABJM theory \cite{Aharony:2008ug} on S$^3$.

\subsection{Solutions with a non-backreacted geometry}
\label{sec:non-backreacted_sol}

Let us start by presenting a class of solutions with a non-backreacted EAdS$_{4}$ geometry and non-trivial profiles for the scalar fields. This is possible in a spacetime with Euclidean signature because $z_{i}$ and $\tilde{z}_{i}$ are independent fields and only the products of $z_{i}$ and $\tilde{z}_{i}$ (and also of their derivatives) couple to the metric $A$ function in the Einstein equation (first equation in (\ref{EOM_4D})). Therefore, non-backreacted solutions exist provided $z_{i} \tilde{z}_{i}=0$ $\forall i$.

The simplest case involves just one non-zero scalar, $z_{i}$ or $\tilde{z}_{i}$, and the rest vanishing. For concreteness, let us consider $z_{1}$ or $\tilde{z}_{1}$ being non-zero and $z_{2}=z_{3}=\tilde{z}_{2}=\tilde{z}_{3}=0$. Within this setup, there is a singular solution given by
\begin{equation}
\label{solution_simplest_singular}
e^{2A} = \sinh^2(g\mu)
\hspace{2mm} , \hspace{2mm}
z_{1} = 0
\hspace{2mm} , \hspace{2mm}
\tilde{z}_{1} = \frac{\tilde{c}_{1}}{\sinh^2\left(\frac{1}{2}g\mu \right)} \ ,
\end{equation}
and a regular solution given by
\begin{equation}
\label{solution_simplest_regular}
e^{2A} = \sinh^2(g\mu)
\hspace{2mm} , \hspace{2mm}
z_{1} = \frac{c_{1}}{\cosh^2\left(\frac{1}{2}g\mu \right)}
\hspace{2mm} , \hspace{2mm}
\tilde{z}_{1} =  0 \ .
\end{equation}
These solutions satisfy the second order equations in (\ref{EOM_4D}) as well as the BPS equations (\ref{BPS_4D}) for the upper choice of sign therein. Therefore, they are supersymmetric solutions. Moreover, if we demand a non-backreacted EAdS$_{4}$ geometry, these two solutions are the only BPS solutions of the system with the two fields $z_{1}$ and $\tilde{z}_{1}$ and the upper sign choice in (\ref{BPS_4D}).

Some comments about the solutions (\ref{solution_simplest_singular})-(\ref{solution_simplest_regular}) are in order. Firstly, the exchange $z_{1} \leftrightarrow \tilde{z}_{1}$ in (\ref{solution_simplest_singular})-(\ref{solution_simplest_regular}) is a symmetry of the second order equations of motion (\ref{EOM_4D}) but not of the BPS equations (\ref{BPS_4D}): the exchange $z_{1} \leftrightarrow \tilde{z}_{1}$ amounts to a different sign choice in the BPS equations (\ref{BPS_4D}) so that the preserved supersymmetries are constructed using either left-invariant or right-invariant Killing spinors on the S$^{3}$. Secondly, solutions involving just one pair of fields $(z_{i},\tilde{z}_{i})$ like, for example, the pair $(z_{1},\tilde{z}_{1})$ in the solutions above, have an enhancement of symmetry from the $\textrm{U}(1)^4$ generic symmetry of the STU-model to an $\textrm{SU}(2)^{4} \sim \textrm{SO}(4) \times \textrm{SO}(4) \subset \textrm{SO}(8)$ symmetry. This $\textrm{SO}(4) \times \textrm{SO}(4)$ symmetry becomes manifest when the solutions are uplifted to backgrounds of 11D supergravity using the formulae in the Appendix~\ref{sec:11D_SO(4)xSO(4)}. The internal geometry is of the form $\textrm{S}^{7}=\mathcal{I} \times \textrm{S}^{3}_{1} \times \textrm{S}^{3}_{2}$ with the two three-spheres $\textrm{S}^{3}_{1,2}$ being responsible for the residual $\textrm{SO}(4) \times \textrm{SO}(4)$ symmetry of the solutions. Thirdly, from (\ref{ds_11D})-(\ref{f_functions}) and (\ref{F4_all})-(\ref{h_functions_uplift}), it becomes clear that $c_{1} \in \mathbb{R}$ renders the 11D metric real and the four-form flux purely imaginary. Also, the unit-disk normalisation condition along the flows requires $|z_{1}(0)| = |c_{1}|<1$.

Since the bulk Lagrangian in (\ref{S_STU-model})-(\ref{V_STU-model}) does not mix the different scalars, the above solutions can be generalised to the STU-model. However, once generalised to the STU-model, they become solutions of the second order equations (\ref{EOM_4D}) but no longer of the BPS equations (\ref{BPS_4D}). Therefore they turn into non-supersymmetric solutions. The singular solution in (\ref{solution_simplest_singular}) is straightforwardly generalised to the STU-model as
\begin{equation}
\label{solution_simplest_singular_general}
e^{2A} = \sinh^2(g\mu)
\hspace{2mm} , \hspace{2mm}
z_{i} = 0
\hspace{2mm} , \hspace{2mm}
\tilde{z}_{i} = \frac{\tilde{c}_{i}}{\sinh^2\left(\frac{1}{2}g\mu \right)}   \ ,
\end{equation}
whereas the regular solution in (\ref{solution_simplest_regular}) has a generalisation of the form
\begin{equation}
\label{solution_simplest_regular_general}
e^{2A} = \sinh^2(g\mu)
\hspace{2mm} , \hspace{2mm}
z_{i} = \frac{c_{i}}{\cosh^2\left(\frac{1}{2}g\mu \right)}
\hspace{2mm} , \hspace{2mm}
\tilde{z}_{i} = 0  \ .
\end{equation}
We will further study the regular solution (\ref{solution_simplest_regular_general}) and compute its gravitational on-shell action and free energy as a function of the parameters $c_{i} \in \mathbb{R}$. Note that the unit-disk normalisation condition along the scalars flow implies $|z_{i}(0)|=|c_{i}| < 1$ $\forall i$. The uplift of the regular solution (\ref{solution_simplest_regular_general}) to eleven-dimensional supergravity requires to generalise the Appendix~\ref{sec:11D_SO(4)xSO(4)} to the full STU-model as done in \cite{Azizi:2016noi}. This goes beyond the scope of this note.

\subsection{Solutions with a backreacted geometry}

Inspired by the Janus solutions of \cite{Anabalon:2022fti}, let us minimally modify the EAdS$_{4}$ geometry in (\ref{EAdS_solution}) and present two new classes of analytic and multi-parametric Euclidean solutions of the STU-model. These two classes of solutions are generically non-supersymmetric although, as we will see, a supersymmetric limit can still be considered.

\subsubsection{Soliton solutions}
\label{sec:solitons}

In the first class of solutions, the spacetime metric (\ref{ds4_ansatz}) is specified by a function
\begin{equation}
\label{solution_soliton_metric}
e^{2A} = \frac{\sinh^2(g\mu)}{k^2} \ ,
\end{equation}
with $k^2>0$ so that the metric is real and Euclidean. Whenever $k^2 \neq 1$, the spacetime metric possesses a singularity at $\mu=0$, as can be seen from the scalar curvature ${R=6 g^2 \left( k^2 - \cosh(2 g \mu) \right) / \sinh^{2}(g \mu)}$. This restricts the range of the solution to $\mu \in \mathbb{R}^{+}$, so we will refer to these solutions as \textit{solitons}.

The metric function in (\ref{solution_soliton_metric}) requires both $z_{i}$ and $\tilde{z}_{i}$ scalars to flow simultaneously. The profiles for these fields are of the form
\begin{equation}
\label{solution_soliton_fields_general}
z_{i} =  \frac{\lambda_{i} \sqrt{k_{i}^2-1}}{k_{i} + \cosh(g\mu)} \hspace{5mm} , \hspace{5mm}
\tilde{z}_{i} = \frac{\lambda_{i}^{-1} \sqrt{k_{i}^2-1}}{k_{i} - \cosh(g\mu)}
 \ ,
\end{equation}
with arbitrary parameters $\lambda_{i}$ and $k_{i}$ subject to the condition
\begin{equation}
\label{k_codition_soliton}
\sum_{i=1}^{3} k_{i}^2=k^2 + 2 \ .
\end{equation}
This class of solutions has a metric singularity at ${\mu=0}$ together with a divergence of the scalars whenever ${k^2_{i} > 1}$. However, such a scalar divergence can be eliminated by taking $k_{i}^2 \le 1$ with the constraint ${\sum k_{i}^2>2}$ so that $k^2>0$ in (\ref{k_codition_soliton}). The unit-disk normalisation condition then fixes $|\lambda_{i}|^{-2}=\frac{1-k_{i}}{1+k_{i}}$ for the scalar flows to reach the boundary of the unit-disk at $\mu=0$, \textit{i.e.} $|z_{i}(0)|=|\tilde{z}_{i}(0)|=1$. These singular solutions are very much alike the Lorentzian flows to Hades in \cite{Anabalon:2022fti}.

Interestingly, when two out of the three $k^2_{i}$ in (\ref{solution_soliton_fields_general}) are set to unity, one is left with a singular but supersymmetric solution with $\textrm{SO}(4) \times \textrm{SO}(4) \subset \textrm{SO}(8)$ residual symmetry. For example, setting $k^2_{2}=k^2_{3}=1$ reduces (\ref{k_codition_soliton}) to $k_{1}^2=k^2$ and yields
\begin{equation}
\label{solution_soliton_susy}
z_{1} =  e^{i \beta} \frac{k+1}{k + \cosh(g\mu)}
\hspace{2mm} , \hspace{2mm}
\tilde{z}_{1} = e^{-i \beta}  \frac{k-1}{k - \cosh(g\mu)} \ ,
\end{equation}
with arbitrary $\beta \in \mathbb{R}$ and $0<k^2<1$. The solution (\ref{solution_soliton_susy}) satisfies the BPS equations (\ref{BPS_4D}). Unfortunately, when uplifted to a background of 11D supergravity using the formulae in the Appendix~\ref{sec:11D_SO(4)xSO(4)}, the singularity at $\mu=0$ persists as it happened for the dual of the Coulomb branch flows investigated in \cite{Anabalon:2022fti}. More concretely, the 11D Ricci scalar goes as $\hat{R} \propto \mu^{-4/3}$ around $\mu=0$. It would be interesting to further investigate the nature of this supersymmetric singularity and its potential implications.

\subsubsection{Wormhole solutions}
\label{sec:wormholes}

The second class of solutions has a regular spacetime metric (\ref{ds4_ansatz}) specified by the metric function 
\begin{equation}
\label{solution_WH_metric}
e^{2A} = \frac{\cosh^2(g\mu)}{k^2} \ ,
\end{equation}
with $k^2>0$ so that the metric is real and Euclidean, and ${\mu \in \mathbb{R}}$. The radius of the $\textrm{S}^{3}$ in the spacetime geometry (\ref{ds4_ansatz}) reaches a minimum size $(gk)^{-1}$ at $\mu=0$ and so we will refer to this second class of solutions as \textit{wormholes}.

The metric function in (\ref{solution_WH_metric}) requires both $z_{i}$ and $\tilde{z}_{i}$ to flow with profiles
\begin{equation}
\label{solution_wormhole_fields_general}
z_{i} = \frac{\lambda_{i} \sqrt{k_{i}^2+1}}{k_{i} - \sinh(g\mu)}
\hspace{5mm} , \hspace{5mm}
\tilde{z}_{i} =  \frac{\lambda^{-1}_{i} \sqrt{k_{i}^2+1}}{k_{i} + \sinh(g\mu)} \ ,
\end{equation}
with arbitrary parameters $\lambda_{i}$ and $k_{i}$ subject to
\begin{equation}
\label{k_codition_wormhole}
\sum_{i=1}^{3} k_{i}^2=k^2 - 2 \ .
\end{equation}
Importantly, when $0< k^2 < 2$, the condition (\ref{k_codition_wormhole}) has solutions of the form $k_{i}= i \, \mathbf{k}_{i}$ with ${\mathbf{k}_{i} \in \mathbb{R}}$. The scalars are non-singular and \textit{necessarily} complex in this case, and read
\begin{equation}
\label{solution_wormhole_fields_general_regular}
z_{i} = \frac{\lambda_{i} \sqrt{1-\mathbf{k}_{i}^2}}{i \, \mathbf{k}_{i} - \sinh(g\mu)}
\hspace{5mm} , \hspace{5mm}
\tilde{z}_{i} =   \frac{\lambda^{-1}_{i} \sqrt{1-\mathbf{k}_{i}^2}}{i \, \mathbf{k}_{i} + \sinh(g\mu)} \ .
\end{equation}
As a result, whenever the real vector $\vec{\mathbf{k}} \equiv (\mathbf{k}_{1},\mathbf{k}_{2},\mathbf{k}_{3})$ (with $\mathbf{k}_{i} \neq 0$ $\forall i$) specifies a point on a two-sphere of radius $(2-k^2) > 0$, the corresponding solution is fully regular. However, the unit-disk parameterisation of the scalars still requires $|z_{i}|,|\tilde{z}_{i}|<1$ all along the scalar flow. Provided this normalisation condition holds, the solution describes an Euclidean wormhole within the STU-model in the $\textrm{SO}(8)$ gauged supergravity. Lastly, the supersymmetric limit of setting two out of the three $\mathbf{k}_{i}^2$ to unity in (\ref{solution_wormhole_fields_general_regular}) is no longer compatible with the condition (\ref{k_codition_wormhole}), which reduces in this case to $-\mathbf{k}_{1}^2=k^2$.
\\

\noindent\textit{\textbf{Example:}} A simple class of regular Euclidean wormholes follows from the identification $\mathbf{k}_{i}^2 = (2-k^2)/3$ $\forall i$. In this case one has
\begin{equation}
\label{modulus_condition_example}
|\lambda_{i}|^{-2} |z_{i}(\mu)|^2  = |\lambda_{i}|^{2} |\tilde{z}_{i}(\mu)|^2  =  \frac{2 (1+k^2)}{(1-2 k^2) + 3 \cosh(2 g\mu)} \ .
\end{equation}
Since $0 < k^2 < 2$, the right-hand side of (\ref{modulus_condition_example}) is regular and has a maximum at $\mu=0$. Then the unit-disk normalisation condition requires
\begin{equation}
\label{modulus_condition_example_normalisation}
\begin{array}{rcrcll}
|z_{i}(0)|^2  & = &  |\lambda_{i}|^{2} \left( \dfrac{3}{2-k^2}-1  \right) & < & 1 & , \\[4mm]
|\tilde{z}_{i}(0)|^2 & = & |\lambda_{i}|^{-2}  \left( \dfrac{3}{2-k^2}-1  \right) & < & 1 & .
\end{array}
\end{equation}
Setting for simplicity $\lambda_{i}=e^{i\beta}$ $\forall i$, with arbitrary $\beta \in \mathbb{R}$, the scalars take the final form
\begin{equation}
\label{solution_wormhole_example}
z_{i} = -\tilde{z}_{i}^{*} =  e^{i \beta} \frac{\sqrt{1+k^2}}{i \sqrt{2-k^2} - \sqrt{3} \sinh(g\mu)} \ ,
\end{equation}
are regular and satisfy $\tilde{z}_{i}=-z_{i}^{*} \in \mathbb{C}$. Moreover, they are compatible with the unit-disk parameterisation condition whenever $0< k^2 < \frac{1}{2}$. Note also that both the \textit{dilatonic} $z_{i}+\tilde{z}_{i}$ and \textit{axionic} $z_{i}-\tilde{z}_{i}$ combinations run non-trivially in this example.

Finally, it is also worth mentioning that solutions with $z_{1}=z_{2}=z_{3}$ (same for $\tilde{z}_{i}$), like the one in (\ref{solution_wormhole_example}), feature an enhancement of symmetry from the generic $\textrm{U}(1)^4$ symmetry of the STU-model to an $\textrm{SU}(3)\times \textrm{U}(1)^2 \subset \textrm{SO}(8)$ symmetry. However, supersymmetry is not preserved in the bulk as the BPS equations (\ref{BPS_4D}) are not satisfied. These solutions are straightforwardly uplifted to Euclidean solutions of eleven-dimensional supergravity using the uplift formulae in the Appendix~\ref{sec:11D_SU(3)xU(1)xU(1)}: the result is a complex metric and a complex (not purely imaginary) four-form flux in 11D.

The complex nature of Euclidean solutions should not be surprising. Metrics with real and imaginary parts are already encountered in the Wick rotation of spinning black holes. Also the Wick rotation of the electrically charged four-dimensional Reissner-Nordstr\"om-AdS black hole has a purely imaginary gauge field that yields a complex metric when uplifted to eleven dimensions. Lastly, black rings do not even have everywhere regular real Euclidean sections \cite{Elvang:2006dd}.

\section{On-shell action and gravitational free energy}

The evaluation of the gravitational on-shell action gives rise to divergences due to the infinite volume in the integral of the bulk action (\ref{S_STU-model}). Such divergences are cured by applying the by now standard procedure of holographic renormalisation \cite{Bianchi:2001kw}. However, the holographic renormalisation procedure comes along with an ambiguity: local finite counter-terms can be added to the renormalised action. This reflects the scheme dependence of the renormalisation procedure in quantum field theory (see \cite{Bianchi:2001de} for the identification of the supersymmetric scheme in a five-dimensional gravitational context). As concluded in \cite{Freedman:2013oja} after the study of the projection of the bulk supersymmetry into the boundary, the identification of sources and VEV's in (\ref{sources_general})-(\ref{vev_general}) requires to include specific cubic counter-terms in the holographic renormalisation procedure. The inclusion of such cubic counter-terms was shown to be crucial to have a matching between the gravitational free energy of the supergravity flows constructed in \cite{Freedman:2013oja} and the free energy of the dual ABJM theory on $\textrm{S}^{3}$ deformed with supersymmetric real mass terms computed using localisation methods.

The multi-parametric families of Euclidean solutions we have constructed in the previous section belong to the same bulk theory as the solutions constructed in \cite{Freedman:2013oja}. Moreover, the solutions here possess a supersymmetric limit whenever two of the scalars $z_{i}$ (same for $\tilde{z}_{i}$) can be smoothly set to vanish upon tuning of the free parameters. This makes us follow the same renormalisation prescription of \cite{Freedman:2013oja} and include the same (finite) cubic counter-terms when renormalising the on-shell action. The result is an on-shell action of the form
\begin{equation}
\label{Renormalised_action}
2\kappa^2  S_{\textrm{on-shell}}  =  2\kappa^2  S_{\textrm{bulk}}+S_{\textrm{GH}}+S_{a}+S_z  \ ,
\end{equation}
with $S_{\textrm{bulk}}$ given in (\ref{S_STU-model}), that contains a set of boundary terms
\begin{equation}
\label{S_extra_terms}
\begin{array}{rcll}
S_{\textrm{GH}}& = & - \displaystyle\int_{\partial M}d^3x\sqrt{h}K \ , \\[4mm]
S_{a}& = & \dfrac{1}{2g} \displaystyle\int_{\partial M}d^3x\sqrt{h} R(h)  \ , \\[4mm]
S_z & = & 2 \, \displaystyle\int_{\partial M}d^3x \, \sqrt{h} \, \left[ e^{\mathcal{K}} \, W \,  \tilde{W} \right]^{\frac{1}{2}}    \\[3mm]
& = & g \displaystyle\int_{\partial M}d^3x \, \sqrt{h} \, \left( 2+ \displaystyle\sum_{i} z_i \tilde z_i \right. \\
& & \left. \hspace{27mm} + \displaystyle\prod_{i} z_{i} + \displaystyle\prod_{i} \tilde{z}_{i} + \ldots\right) \ .
\end{array}
\end{equation}
The above boundary terms include the standard counter-terms that follow from the near-boundary analysis of a generic supergravity solution (see \textit{e.g.} \cite{Bianchi:2001kw}), the specific cubic counter-term discussed in \cite{Freedman:2013oja} (\textit{i.e.} last line of (\ref{S_extra_terms})), and additional terms -- denoted by the ellipsis -- which vanish at the boundary when taking the limit $|\mu|\rightarrow \infty$. The quantity $K$ entering $S_{\textrm{GH}}$ is the trace of the extrinsic curvature defined as $K_{a b}=\nabla_{(a} N_{b)}=\frac{1}{2}\mathcal{L}_{N}h_{ab}= \pm \frac{1}{2} h_{ab}'$. Here $N_{a}=\pm \delta_{a}^{\mu}$ is the vector normal to the boundary $\partial M$ and $h_{a b}=g_{a b}-N_{a}N_{b}$ is the induced metric \footnote{The ambiguity of the sign in $N_{a}$ is because, in the wormhole solution, the boundary consists of two disconnected pieces located at $\mu = \pm \infty$.}. The quantity $R(h)$ entering $S_{a}$ is the Ricci scalar of this metric. Finally, the functions $\mathcal{K}$, $W$ and $\tilde{W}$ are the K\"ahler potential and superpotentials given in (\ref{Kahler_func}) and (\ref{W_funcs}). We note in passing that the cubic counter-terms $z_{1}z_{2}z_{3}$ and $\tilde{z}_{1}\tilde{z}_{2}\tilde{z}_{3}$ in the last line of (\ref{S_extra_terms}) break the $\mathbb{C}_{i}$ scaling symmetries of the bulk action (\ref{S_STU-model}).

Starting from the on-shell action (\ref{Renormalised_action}), the computation of the gravitational free energy, $J_{\textrm{on-shell}}$, requires a precise identification of sources and VEVs from the scalar expansions (\ref{scalars_expasions_ab}) around the boundary. The reason why is that a Legendre transformation of $S_{\textrm{on-shell}}$ might be needed for $J_{\textrm{on-shell}}$ to depend on the particular combinations of leading and sub-leading coefficients in (\ref{scalars_expasions_ab}) specifying the sources in the dual field theory. With the identification of sources given in (\ref{sources_general}), the gravitational free energy takes the form
\begin{equation}
\label{F_general_expression}
J_{\textrm{on-shell}} = S_{\textrm{on-shell}} + \frac{1}{2 \kappa^2} \Delta S \ ,
\end{equation}
with
\begin{equation}
\label{DeltaS_contribution}
\Delta S = - \frac{1}{2} \displaystyle \sum_{i} \int_{\textrm{S}^3} dx^3 (a_{i} + \tilde{a}_{i}) \left(  \frac{\delta S_{\textrm{on-shell}}}{\delta a_{i}} + \frac{\delta S_{\textrm{on-shell}}}{\delta \tilde{a}_{i}}   \right) \ .
\end{equation}
And it is precisely at this point where the finite cubic counter-terms entering $S_{z}$ in (\ref{S_extra_terms}) play a crucial role as they guarantee that
$\delta_{a_{i}} S_{\textrm{on-shell}} + \delta_{\tilde{a}_{i}} S_{\textrm{on-shell}} \propto {(\tilde{b}_{i} - a_{1}a_{2}a_{3}/a_{i}) + (b_{i} - \tilde{a}_{1}\tilde{a}_{2}\tilde{a}_{3}/\tilde{a}_{i})}$ and, therefore, the gravitational free energy $J_{\textrm{on-shell}}$ is a function of the sources in (\ref{sources_general}).

\subsection{Non-backreacted EAdS$_{4}$ solutions}

Let us recall the non-backreacted regular solution in (\ref{solution_simplest_regular_general}), namely,
\begin{equation}
\label{solution_simplest_regular_general_2}
e^{2A} = \sinh^2(g\mu)
\hspace{2mm} , \hspace{2mm}
z_{i} = \frac{c_{i}}{\cosh^2\left(\frac{1}{2}g\mu \right)}
\hspace{2mm} , \hspace{2mm}
\tilde{z}_{i} = 0  \ .
\end{equation}
When reaching the boundary at ${\mu = \infty}$, the scalar fields feature the asymptotic expansions
\begin{equation}
\label{scalars_expasions_non-backreacted}
z_{i} =  - 4 c_{i} \displaystyle\sum_{n=1}^{\infty} (-1)^{n} \, n \, e^{-n g \mu}
\hspace{5mm} \textrm{ and } \hspace{5mm}
\tilde{z}_{i} = 0 \ .
\end{equation}
A direct comparison between (\ref{scalars_expasions_ab})-(\ref{vev_general}) and (\ref{scalars_expasions_non-backreacted}) shows that
\begin{equation}
\begin{array}{rcl}
a_{i} \mp \tilde{a}_{i} &=& 4 c_{i} \ , \\[2mm]
\left( b_{i} - \frac{\tilde{a}_{1}\tilde{a}_{2}\tilde{a}_{3}}{\tilde{a}_{i}} \right) \pm \left( \tilde{b}_{i} -\frac{a_{1}a_{2}a_{3}}{a_{i}} \right) &=& - 16 \left(  \frac{c_{i}}{2} \pm \frac{c_{1}c_{2}c_{3}}{c_{i}}\right) \ ,
\end{array}
\end{equation}
and thus the sources in (\ref{sources_general}) and VEVs in (\ref{vev_general}) are generically activated if $c_{i} \neq 0$. The evaluation of the on-shell action (\ref{Renormalised_action}) on the solution (\ref{solution_simplest_regular_general_2}) gives
\begin{equation}
\label{S_on-shell_non-backreacted}
S_{\textrm{on-shell}} \Big|_{\textrm{non-back}}  = S^{\textrm{EAdS}_{4}}_{\textrm{on-shell}} \left(1+4\prod_{i}c_{i}\right) \ ,
\end{equation}
where
\begin{equation}
S^{\textrm{EAdS}_{4}}_{\textrm{on-shell}} = \frac{2\pi^2}{\kappa^2  g^2} \ ,
\end{equation}
is the on-shell action of the EAdS$_{4}$ solution in (\ref{EAdS_solution}). The first contribution to the r.h.s of (\ref{S_on-shell_non-backreacted}) comes from the bulk term in (\ref{Renormalised_action}) whereas the second $c_{i}$-dependent contribution originates from the non-zero cubic counter-terms in $S_{z}$ of (\ref{S_extra_terms}). Without the cubic counter-terms in the holographic renormalisation procedure, the on-shell action of the solution (\ref{solution_simplest_regular_general_2}) would be the same as that of EAdS$_{4}$. Also, from (\ref{S_on-shell_non-backreacted}), it follows that the on-shell action of the non-backreacted solution (\ref{solution_simplest_regular_general_2}) is the same as that of EAdS$_{4}$ whenever (at least) one of the scalars $z_{i}$ vanishes.

The gravitational free energy of the solution (\ref{solution_simplest_regular_general_2}) is given by (\ref{F_general_expression}) and (\ref{DeltaS_contribution}) with
\begin{equation}
\label{DeltaS_non_backreacted}
\begin{array}{rcl}
\dfrac{\delta S_{\textrm{on-shell}}}{\delta a_{i}} &=&  - \dfrac{\sqrt{g_{3}}}{16 g^2 \kappa^2} \left(  \tilde{b}_{i} - \dfrac{a_{1}a_{2}a_{3}}{a_{i}}\right) \ , \\[4mm]
\dfrac{\delta S_{\textrm{on-shell}}}{\delta \tilde{a}_{i}}   & = & -
\dfrac{\sqrt{g_{3}}}{16 g^2 \kappa^2} \left(  b_{i} - \dfrac{\tilde{a}_{1}\tilde{a}_{2}\tilde{a}_{3}}{\tilde{a}_{i}}\right) \ ,
\end{array}
\end{equation}
where $g_{3}$ is the determinant of the round metric on the three-sphere $\textrm{S}^{3}$ of unit radius. Substituting the $a_{i}$, $\tilde{a}_{i}$, $b_{i}$ and $\tilde{b}_{i}$ coefficients that follow from direct comparison between the general expansions in (\ref{scalars_expasions_ab})-(\ref{vev_general}) and the ones in (\ref{scalars_expasions_non-backreacted}), the resulting gravitational free energy is given by
\begin{equation}
\label{J_on-shell_non-backreacted} 
J_{\textrm{on-shell}}\Big|_{\textrm{non-back}} = 
J^{\textrm{EAdS}_{4}}_{\textrm{on-shell}} 
\left(1-\sum_{i}\mathcal{C}_{i}\right) \ ,
\end{equation}
where
\begin{equation}
\label{MathcalC_def}
\mathcal{C}_{i} = c_{i}^2  + \tfrac{2}{3} \,  c_{1} c_{2} c_{3} \ ,
\end{equation}
and
\begin{equation}
\label{J_EAdS}
\begin{array}{rcl}
J^{\textrm{EAdS}_{4}}_{\textrm{on-shell}}    &=& \dfrac{2\pi^2}{\kappa^2  g^2} \ ,
\end{array}
\end{equation}
is the universal gravitational free energy of the EAdS$_{4}$ vacuum in (\ref{EAdS_solution}). As a consequence of $F$-extremisation, the gravitational free energy in (\ref{J_on-shell_non-backreacted}) is maximised at the conformal case, \textit{i.e.} $c_{i}=0$ $\forall i$, so that
\begin{equation}
J_{\textrm{on-shell}}\Big|_{\textrm{non-back}} <
J^{\textrm{EAdS}_{4}}_{\textrm{on-shell}} \ ,
\end{equation}
provided the unit-disk condition $|c_{i}|<1$ holds.

Let us comment on the gravitational free energy we have obtained. For generic values of the parameters $c_{i}$, the solution (\ref{solution_simplest_regular_general_2}) is non-supersymmetric and, in principle, one could consider a renormalisation scheme different from the supersymmetric one. For example, if applying the minimal subtraction scheme that follows from the near-boundary analysis -- so no finite cubic counter-terms are included -- the gravitational free energy changes to $J^{\textrm{EAdS}_{4}}_{\textrm{on-shell}} 
\left(1-\sum_{i}c_{i}^2\right)$ with no mixing between the $z_{i}$'s. Positivity of this free energy further restricts the range of the parameters $c_{i}$ to lie inside the unit radius $c_{i}$-ball defined by the condition $\sum_{i}c_{i}^2<1$. In this note, and in order to make contact with \cite{Freedman:2013oja} in the supersymmetric case, we have applied the supersymmetric scheme and, therefore, the gravitational free energy in (\ref{J_on-shell_non-backreacted}) depends on the finite cubic counter-terms in $S_{z}$ of (\ref{S_extra_terms}): the cubic term in the r.h.s of (\ref{MathcalC_def}) stems from such finite counter-terms. Positivity of (\ref{J_on-shell_non-backreacted}) then requires the parameters $c_{i}$ to lie inside a deformed $c_{i}$-ball with four lumps defined by the cubic condition $\sum_{i}c_{i}^2 + 2 \, c_{1}c_{2}c_{3}<1$.

\subsubsection{Supersymmetric limit and holography}

Setting two $c_{i}$ to zero in the solution (\ref{solution_simplest_regular_general_2}), \textit{i.e.} ${c_{2}=c_{3}=0}$, reduces it to the supersymmetric and $\textrm{SO}(4)\times \textrm{SO}(4) \sim \textrm{SU}(2)^4$ invariant solution in (\ref{solution_simplest_regular}), namely,
\begin{equation}
\label{solution_simplest_regular_2}
e^{2A} = \sinh^2(g\mu)
\hspace{2mm} , \hspace{2mm}
z_{1} = \frac{c_{1}}{\cosh^2\left(\frac{1}{2}g\mu \right)}
\hspace{2mm} , \hspace{2mm}
\tilde{z}_{1} =  0 \ .
\end{equation}
The general gravitational free energy in (\ref{J_on-shell_non-backreacted}) reduces in this case to
\begin{equation}
\label{J_on-shell_non-backreacted_susy}
J_{\textrm{on-shell}} \Big|_{\textrm{SO}(4)\times \textrm{SO}(4)} = J^{\textrm{EAdS}_{4}}_{\textrm{on-shell}} \left(1-c_{1}^{2}\right) \ .
\end{equation}
The solution in (\ref{solution_simplest_regular_2}) lies precisely at the intersection between the general class of non-supersymmetric solutions in (\ref{solution_simplest_regular_general_2}) and the class of ${\mathcal{N}=2}$ flows put forward in \cite{Freedman:2013oja} describing the holography of
$F$-maximisation \cite{Jafferis:2010un}. The uplift of (\ref{solution_simplest_regular_2}) to 11D supergravity was performed in \cite{Gautason:2023igo} and agrees with the general uplift formulae for the $\textrm{SO}(4) \times \textrm{SO}(4)$ invariant sector collected in Appendix~\ref{sec:11D_SO(4)xSO(4)}.

Following \cite{Freedman:2013oja}, let us look at the field theory dual of the solution (\ref{solution_simplest_regular_2}). This is the ABJM superconformal field theory (SCFT) of \cite{Aharony:2008ug} placed on (a unit radius) S$^3$ and deformed with a specific supersymmetric real mass parameter $\delta_{1}$. In general, there are three such real mass parameters $\delta_{i}$ ($i=1,2,3$) compatible with $\mathcal{N}=2$ supersymmetry which modify the assignment of $\textrm{U}(1)_{\textrm{R}}$ R-charges. The real mass parameters $\delta_{i}$ break conformality and also the $\textrm{U}(1)_{\textrm{R}} \times \textrm{SU}(2) \times \textrm{SU}(2) \times \textrm{U}(1)_{b} \subset \textrm{SO}(8)_{\textrm{R}}$ symmetry manifest in the $\mathcal{N}=2$ superfield formulation of the (undeformed) ABJM SCFT down to its Cartan subgroup $\textrm{U}(1)_{\textrm{R}} \times \textrm{U}(1) \times \textrm{U}(1) \times \textrm{U}(1)_{b} \subset \textrm{SO}(8)_{\textrm{R}}$.

In the $\mathcal{N}=2$ formulation of \cite{Aharony:2008ug}, ABJM theory is a SCFT with gauge group $\textrm{U}(N) \times \textrm{U}(N)$ and Chern--Simons (CS) levels $(k,-k)$ -- in our case $k=1$ and the internal space in the gravity side is the round S$^{7}$ --. The theory consists of two vectors multiplets $(A,\sigma,\lambda,D)$ and $(\tilde{A},\tilde{\sigma},\tilde{\lambda},\tilde{D})$ for the $\textrm{U}(N) \times \textrm{U}(N)$ gauge group together with four chiral matter multiplets $(Z^{a},\chi^{a},F^{a})$ and $(W_{a},\eta_{a},G_{a})$, with $a=1,2$, transforming in the $(\bf{\bar{N}},\bf{N})$ and $(\bf{N},\bf{\bar{N}})$ representations of the gauge group. There is also a quartic superpotential for the matter multiplets which has R-charge $2$ and is invariant under the $\textrm{SU}(2) \times \textrm{SU}(2)\times \textrm{U}(1)_{b}$ flavour symmetry. It is given by
\begin{equation}
\label{W_ABJM}
W \propto \textrm{Tr} \left(  \epsilon_{ab}\epsilon^{cd} Z^{a} W_{c} Z^{b} W_{d} \right) \ .
\end{equation}
Particularising the Lagrangian in \cite{Freedman:2013oja} to describe, holographically, the solution in (\ref{solution_simplest_regular_2}) one finds
\begin{equation}
\label{Lagrangian_delta1}
\mathcal{L} = \mathcal{L}_{\textrm{SCFT}} + \delta_{1} \left[ \mathcal{O}^{1}_{B} - \delta_{1} \mathcal{O}_{S} \right] + \delta_{1} \mathcal{O}^{1}_{F} \ ,
\end{equation}
where $\mathcal{L}_{\textrm{SCFT}}$ is the ABJM SCFT Lagrangian and $\delta_{1}$ is a specific supersymmetric real mass parameter that turns on bosonic and fermionic operators. In particular, it turns on the bosonic $\mathcal{O}^{1}_{B}$ and fermionic $\mathcal{O}^{1}_{F}$ operators
\begin{equation}
\label{Operators_B_F}
\begin{array}{rcl}
\mathcal{O}^{1}_{B} &=& \textrm{tr}\left[ Z^{\dagger}_{a}Z^{a} -  W^{\dagger a}W_{a} \right] \ , \\[4mm]
\mathcal{O}^{1}_{F} & = & \textrm{tr}\left[ \chi^{\dagger a} \chi_{a}  - \eta^{\dagger}_{a} \eta^{a}\right] + 2 i (\sigma-\tilde{\sigma})\mathcal{O}_{S} \ ,
\end{array}
\end{equation}
dual to the proper-scalar $z_{1}+\tilde{z}_{1}$ and the pseudo-scalar $z_{1}-\tilde{z}_{1}$ in the $\textrm{SO}(8)$ supergravity, together with an additional Konishi-like bosonic operator
\begin{equation}
\label{Operator_S}
\mathcal{O}_{S} = \textrm{tr}\left[ Z^{\dagger}_{a}Z^{a} + W^{\dagger a}W_{a} \right] \ ,
\end{equation}
which does not have an associated scalar in the $\textrm{SO}(8)$ supergravity \footnote{Spinless fields in the $\textrm{SO}(8)$ supergravity transform in the $\bf{35}_{v}$ (proper-scalars) and $\bf{35}_{c}$ (pseudo-scalars) symmetric and \textit{traceless} representations of $\textrm{SO}(8)$. The bosonic operator $\mathcal{O}_{S}$ in (\ref{Operator_S}) corresponds to the trace of the boson bilinears and, therefore, has no associated scalar in the $\textrm{SO}(8)$ supergravity.}. Remarkably, amongst the operators with a dual spinless field in the $\textrm{SO}(8)$ supergravity, those in (\ref{Operators_B_F}) are the only two singlets under the $\textrm{SU}(2) \times \textrm{SU}(2) \times \textrm{U}(1)_{b}$ flavour symmetry of the ABJM SCFT in the $\mathcal{N}=2$ formulation: under this symmetry the chiral multiplets $(Z^{a},\chi^{a},F^{a})$ and $(W_{a},\eta_{a},G_{a})$ transform in the ${(\bf{2},\bf{1})}_{1}$ and ${(\bf{1},\bf{\bar{2}})}_{-1}$ representations, respectively \footnote{See Table~1 of \cite{Freedman:2013oja} for a summary of the $\textrm{U}(1)_{\textrm{R}} \times \textrm{SU}(2) \times \textrm{SU}(2) \times \textrm{U}(1)_{b}$ charges of the various fields in the ABJM SCFT}.

The parameter $\delta_{1}$ in (\ref{Lagrangian_delta1}) determines the assignment of $\textrm{U}(1)_{\textrm{R}}$ R-charges for the scalar component of the chiral superfields to be
\begin{equation}
\label{R-charges_SO(4)}
R[Z^{1}]=R[Z^{2}]=\tfrac{1}{2}+\delta_{1} 
\, , \,
R[W_{1}]=R[W_{2}]=\tfrac{1}{2}-\delta_{1}  \ .
\end{equation}
On the other hand, the free energy of the ABJM theory on $\textrm{S}^3$ deformed with $\mathcal{N}=2$ supersymmetric real mass terms can be obtained as a function of a set of trial R-charges using localisation methods \cite{Kapustin:2009kz,Jafferis:2010un}. To leading order in $N$ and for CS level $k=1$ the result is \cite{Jafferis:2011zi}
\begin{equation}
\label{F_ABFM}
\mathcal{F} = \mathcal{F}_{0} \sqrt{16 \, R[Z^{1}] \, R[Z^{2}] \, R[W_{1}] \, R[W_{2}]} \ ,
\end{equation} 
with $\mathcal{F}_{0} = \sqrt{2} \pi N^{3/2} / 3$ and where the $R$'s are the trial charges of the four chiral matter fields in the theory. The superpotential of the undeformed ABJM SCFT theory (\ref{W_ABJM}) is quartic on the matter fields and has R-charge $2$. This implies that $R[Z^{1}] + R[Z^{2}] + R[W_{1}] + R[W_{2}]=2$, rule that is obeyed by the R-charges in (\ref{R-charges_SO(4)}). A direct substitution of the R-charges (\ref{R-charges_SO(4)}) into (\ref{F_ABFM}) yields
\begin{equation}
\label{F_QFT_SO(4)xSO(4)}
\mathcal{F}_{\textrm{SO}(4) \times \textrm{SO}(4)} =  \mathcal{F}_{0} \, (1 - (2\delta_{1})^2 ) \ ,
\end{equation}
which matches the gravitational free energy in (\ref{J_on-shell_non-backreacted_susy}) for the supersymmetric $\textrm{SO}(4) \times \textrm{SO}(4)$ invariant solution (\ref{solution_simplest_regular_2}) provided 
\begin{equation}
\mathcal{F}_{0} = J^{\textrm{EAdS}_{4}}_{\textrm{on-shell}} \ ,
\end{equation}
and the identification ${c_{1} = 2 \delta_{1}}$. The extremisation of the free energy in (\ref{F_QFT_SO(4)xSO(4)}) fixes $\delta_{1}=0$ and assigns canonical R-charges of $\frac{1}{2}$ to the four chiral fields in (\ref{R-charges_SO(4)}). The free energy in (\ref{F_QFT_SO(4)xSO(4)}) then reduces to $\mathcal{F}_{0}$ which, using the holographic mapping between $\kappa^2$ and $N$ in the ABJM SCFT at CS level $k=1$, matches the gravitational free energy of the EAdS$_{4}$ vacuum in (\ref{J_EAdS}).

\subsection{Backreacted wormhole solutions}

Let us now investigate the backreacted solution in (\ref{solution_WH_metric}) and (\ref{solution_wormhole_fields_general_regular}). As we demonstrated with the example of Section~\ref{sec:wormholes}, the various parameters in the solution can be chosen such that the solution is regular in the domain $\mu \in (-\infty,\infty)$, and the scalars stay within the unit-disk all along the flow. Having one of these regular and classically well-defined solutions in mind, we will proceed an compute its on-shell action.

The boundary of spacetime now consists of two pieces located at $\mu = \pm \infty$. The asymptotic expansions of the scalars in (\ref{solution_wormhole_fields_general_regular}) around $\mu = \pm \infty$ yield
\begin{equation}
\label{scalars_expasions_backreacted}
\begin{array}{lll}
z_{i} &=&  2 \, \lambda_{i} \, \sqrt{1-\mathbf{k}^2_{i}} \, \big( \pm e^{-g |\mu|} - 2 \, i \, \mathbf{k}_{i} \,  e^{-2 g |\mu|} + \ldots \big) \ , \\[4mm]
\tilde{z}_{i} &=& 2 \, \lambda_{i}^{-1} \, \sqrt{1-\mathbf{k}^2_{i}} \, \big( \mp e^{- g |\mu|} - 2 \, i \, \mathbf{k}_{i} \,  e^{- 2 g |\mu|} + \ldots \big) \ ,
\end{array}
\end{equation}
where the upper (lower) sign corresponds to the piece of the boundary at $\mu = + \infty$ ($\mu = - \infty$). Using the asymptotic expansions in (\ref{scalars_expasions_backreacted}) to evaluate the boundary terms, the on-shell action (\ref{Renormalised_action}) for the wormholes takes the form
\begin{equation}
S_{\textrm{on-shell}} \Big|_{\textrm{WH}} = S_{\textrm{cubic}} \Big|_{\partial M_{+}} + S_{\textrm{cubic}} \Big|_{\partial M_{-}} =  0 \ .
\end{equation}
The finite contributions coming from the cubic counter-terms in $S_{z}$ of (\ref{S_extra_terms}) at the two pieces $\partial M_{\pm}$ (respectively at $\mu = \pm \infty$) of the boundary read
\begin{equation}
\label{S_cubic_wormholes}
S_{\textrm{cubic}} \Big|_{\partial M_{\pm}} = \mp \frac{2 \pi^2}{g^2 k^3} \left(  \Lambda - \frac{1}{\Lambda}\right) \displaystyle\prod_{i} \sqrt{1-\mathbf{k}_{i}^2}  \ ,
\end{equation}
with $\Lambda \equiv \prod_{j} \lambda_{j}$. Except for the finite contributions in (\ref{S_cubic_wormholes}), all the other contributions to the on-shell action (\ref{Renormalised_action}) cancel separately on each piece of the boundary of the wormhole.

Using the boundary expansions for the scalar fields in (\ref{scalars_expasions_backreacted}), one finds that 
\begin{equation}
\label{DeltaS_WH}
\begin{array}{rcl}
\dfrac{\delta S_{\textrm{on-shell}}}{\delta a_{i}} &=&  - \dfrac{\sqrt{g_{3}}}{16 g^2 \kappa^2 k^{3}} \left(  \tilde{b}_{i} - \dfrac{a_{1}a_{2}a_{3}}{a_{i}}\right) \ , \\[4mm]
\dfrac{\delta S_{\textrm{on-shell}}}{\delta \tilde{a}_{i}}   & = & -
\dfrac{\sqrt{g_{3}}}{16 g^2 \kappa^2 k^{3}} \left(  b_{i} - \dfrac{\tilde{a}_{1}\tilde{a}_{2}\tilde{a}_{3}}{\tilde{a}_{i}}\right) \ ,
\end{array}
\end{equation}
and an explicit computation of (\ref{DeltaS_contribution}) for the wormholes in (\ref{solution_WH_metric}) and (\ref{solution_wormhole_fields_general_regular}) yields $\Delta S |_{\partial M_{+}} = - \Delta S |_{\partial M_{-}}$. As a result, the total gravitational free energy vanishes again due to an exact cancellation between the two pieces of the wormhole boundary. Namely,
\begin{equation}
J_{\textrm{on-shell}} \Big|_{\textrm{WH}} = J_{\textrm{on-shell}} \Big|_{\partial M_{+}} + J_{\textrm{on-shell}} \Big|_{\partial M_{-}} =  0 \ .
\end{equation}
Importantly, this result is independent of the holographic renormalisation scheme, as any other choice of cubic counter-terms would have also resulted in a null contribution to the gravitational free energy -- and also to the on-shell action -- due to the antisymmetry of the scalar boundary conditions (\ref{scalars_expasions_backreacted}) at $\mu=\pm \infty$.

\section{Conclusions}

In this note we have constructed new classes of $\textrm{U}(1)^4$-invariant Euclidean solutions in the four-dimensional STU-model (with vectors turned off) that arises upon compactification of 11D supergravity on S$^7$. Together with some potentially pathological examples featuring spacetime singularities -- like the solitons in Section~\ref{sec:solitons}-- we presented two classes of fully regular multi-parametric Euclidean solutions. 

The first class describes flows with running scalars $z_{i}$ ($i=1,2,3$) but a non-backreacted EAdS$_{4}$ spacetime. This is possible in Euclidean signature, where the scalars $z_{i}$ and their would-be conjugates $\tilde{z}_{i}$ are independent fields so that a non-trivial solution with $z_{i} \tilde{z}_{i}=0$ $\forall i$ has a vanishing energy-momentum tensor. We constructed the three-parameter family of non-supersymmetric solutions in (\ref{solution_simplest_regular_general}) and computed its on-shell action and gravitational free energy employing holographic renormalisation methods. For the particular choice of parameters $c_{i}\neq 0$ and $c_{j}=c_{k}=0$ in (\ref{solution_simplest_regular_general}), with $i \neq j \neq k$, the solution features an $\textrm{SO}(4) \times \textrm{SO}(4)$ symmetry enhancement, becomes supersymmetric, and belongs to the class of flows put forward in \cite{Freedman:2013oja}. In this supersymmetric limit, the gravitational free energy was shown to match that of ABJM placed on S$^3$ and deformed with a specific real mass term which, in the $\mathcal{N}=2$ formulation of the theory, preserves the full $\textrm{SU}(2) \times \textrm{SU}(2)\times \textrm{U}(1)_{b}$ flavour symmetry group. The embedding of this supersymmetric solution into 11D supergravity showed that, when the profile for the running scalar is taken to be real, \textit{i.e.} $c_{i} \in \mathbb{R}$, the uplift to 11D produces a real metric and a purely imaginary four-form flux. This agrees with the standard statement that ``axions" flip the sign of their kinetic term in the Euclidean theory.

The second class describes regular wormholes with scalar profiles satisfying $z_{i} \tilde{z}_{i} \neq 0$ and supporting the wormhole geometry. We constructed the multi-parametric family of wormholes in (\ref{solution_WH_metric}) and (\ref{solution_wormhole_fields_general_regular}), and showed that a supersymmetric limit is not possible for these solutions. We also computed the on-shell action and gravitational free energy of the wormholes and found that the latter always vanishes irrespectively of supersymmetry and the holographic renormalisation scheme. It would be interesting to understand this result from a field-theoretic perspective. However, the absence of supersymmetry makes a holographic test of the zero gravitational free energy difficult.

Unlike for the non-backreacted flow solutions, the scalars $z_{i}$ and $\tilde{z}_{i}$ are necessarily complex in the wormholes. In the particular example with $\textrm{SU}(3)\times \textrm{U}(1)^2$ symmetry presented in Section~\ref{sec:wormholes}, it happens that $\tilde{z}_{i} =- z_{i}^{*}\in \mathbb{C}$ (\textit{not} $\tilde{z}_{i}=z_{i}^{*}$), but each pair $(z_{i},\tilde{z}_{i})$ still comprises two real degrees of freedom. In addition, the a priori \textit{axionic} combination $z_{i}-\tilde{z}_{i}$ has a non-trivial profile providing an \textit{axionic} charge to the wormhole. Upon uplift of this example to eleven dimensions, both the 11D metric and the four-form flux turn out to have non-trivial real and imaginary parts.

It is also interesting to contrast the class of wormholes presented in this note with the Maldacena--Maoz construction of Euclidean wormholes \cite{Maldacena:2004rf}. In the latter, the foliation of EAdS space is additionally compactified in order to generate the Euclidean wormhole. This compactification procedure imposes identifications that break supersymmetry even asymptotically. In our case, and despite being non-supersymmetric, the wormholes asymptote supersymmetric EAdS vacua on both sides of the solution, thus ensuring a theory with a stable ground state. It would be interesting to further investigate the stability of the wormholes presented here as part of the swampland program, and to assess their allowability in light of the recent discussion in \cite{Witten:2021nzp}. Also their fate in semi-classical (super)gravity -- as an approximation to quantum gravity -- deserves further investigations.

Finally, the class of wormholes presented in this note was inspired by (the Euclidean continuation of) the Lorentzian Janus solutions put forward in \cite{Anabalon:2022fti}. Supersymmetric Lorentzian Janus solutions with identifications that make them wormhole-like (\textit{e.g.} with non-trivial first homotopy group) have been constructed in \cite{Anabalon:2018rzq, Anabalon:2020loe} using vector fields instead of scalars. Their generalisation to include non-trivial running scalars should be possible and could serve as a starting point to construct analytic families of supersymmetric Euclidean wormholes in M-theory. We hope to come back to this and related issues in the future.
\\

\noindent\textbf{Acknowledgements}:  
We are grateful to C. Hoyos, D. Rodr\'iguez-G\'omez and P. Soler for interesting conversations. The work of AA is supported in part by the FAPESP visiting researcher award 2022/11765-7 and the FONDECYT grants 1200986, 1210635, 1221504 and 1230853. The work of AG is supported by the Spanish government grant PGC2018-096894-B-100 and by the Principado de Asturias through the grant FC-GRUPIN-IDI/2018/000174.

\appendix

\section{Uplifts to 11D supergravity}
\label{app:appendix_11D}

The bosonic sector of eleven-dimensional supergravity consists of a metric field $\hat{G}_{MN}$ and a three-form potential $\hat{A}_{(3)}$ with field strength $\hat{F}_{(4)}=d\hat{A}_{(3)}$. The Euclidean equations of motion read \cite{Nakayama:2016xzs}
\begin{equation}
\label{EOM_11D_sugra_Euclidean}
\begin{array}
[c]{rll}%
d(*_{11} \hat{F}_{(4)}) + \frac{i}{2} \, \hat{F}_{(4)} \wedge\hat{F}_{(4)} & = & 0 \ ,\\[2mm]
\hat{R}_{MN} - \frac{1}{12} \left(  (\hat{F}\hat{F})_{MN} - \frac{1}{12} \, (\hat{F}\hat{F}) \, \hat{G}_{MN} \right)  & = & 0 \ ,
\end{array}
\end{equation}
where we have denoted $(\hat{F}\hat{F})_{MN} \equiv \hat{F}_{MPQR} \, \hat{F}_{N}{}^{PQR}$ and $(\hat{F}\hat{F}) \equiv \hat{F}_{PQRS} \, \hat{F}^{PQRS}$. In addition, the field strength $\hat{F}_{(4)}$ satisfies the source-less Bianchi identity
\begin{equation}
\label{BI_11D_sugra}
d\hat{F}_{(4)} = 0 \ .   
\end{equation}

In this appendix we present the necessary formulae to uplift the simplest four-dimensional solutions discussed in the main text to Euclidean solutions of eleven-dimensional supergravity. More concretely, we present the embedding of the $\textrm{SO}(4) \times \textrm{SO}(4)$ and $\textrm{SU}(3) \times \textrm{U}(1)^2$ invariant sectors of the maximal SO(8) gauged supergravity into 11D supergravity.

\subsection{$\textrm{SO}(4) \times \textrm{SO}(4)$ invariant sector}
\label{sec:11D_SO(4)xSO(4)}

When only one pair of scalars $(z_{i},\tilde{z}_{i})$ is non-zero -- we will choose $(z,\tilde{z}) \equiv (z_{1},\tilde{z}_{1})$ without loss of generality -- the corresponding solutions belong to the $\textrm{SO}(4) \times \textrm{SO}(4)$ invariant sector of the maximal $\textrm{SO}(8)$ supergravity. The uplift of this sector to Lorentzian supergravity in eleven dimensions can be found in \cite{Bobev:2013yra}. Here we extend the uplift formulae therein to Euclidean signature and adapt it to the unit-disk parameterisation of the scalar fields $z$ and $\tilde{z}$ we are using in this work.

The 11D (warped) geometry is of the form $\mathcal{M}_{4} \times \textrm{S}^{7}$ with $\textrm{S}^{7}=\mathcal{I} \times \textrm{S}^{3}_{1} \times \textrm{S}^{3}_{2}$. The two three-spheres $\textrm{S}^{3}_{1,2}$ in the internal geometry are responsible for the residual ${\textrm{SO}(4) \times \textrm{SO}(4)}$ symmetry of the solutions. Denoting ${\alpha \in [0,\frac{\pi}{2}]}$ the coordinate along the interval $\mathcal{I}$, the 11D metric is given by
\begin{equation}
\label{ds_11D}
ds_{11}^2 = \Delta^2 \left[  ds_{4}^2 + \frac{4}{g^2} \left(   d\alpha^2 + \frac{\cos^2\alpha}{f_{1}} ds^2_{\textrm{S}^3_{1}} + \frac{\sin^2\alpha}{f_{2}} ds^2_{\textrm{S}^3_{2}}  \right) \right]  ,
\end{equation}
where the external spacetime metric $ds_{4}^2$ is given in (\ref{ds4_ansatz}) and
\begin{equation}
\label{ds_three-spheres}
ds^2_{\textrm{S}^3_{1}} = \tfrac{1}{4} \left(  \sigma_{1}^2 + \sigma_{2}^2 + \sigma_{3}^2 \right)
\hspace{2mm} , \hspace{2mm}
ds^2_{\textrm{S}^3_{2}} = \tfrac{1}{4} \left(  \hat{\sigma}_{1}^2 + \hat{\sigma}_{2}^2 + \hat{\sigma}_{3}^2 \right) \ ,
\end{equation}
denote the line elements on the two internal three-spheres. The warping factor in (\ref{ds_11D}) takes the form
\begin{equation}
\label{Delta_function}
\Delta^2 = f_{1}^{\frac{1}{3}} \, f_{2}^{\frac{1}{3}} \ ,  
\end{equation}
in terms of two functions
\begin{equation}
\begin{array}{rcll}
\label{f_functions}
f_{1} &=& \cos^2\alpha \,  \dfrac{(1+z)(1+\tilde{z})}{1-z \, \tilde{z}}  + \sin^2\alpha & , \\[4mm]
f_{2} &=& \sin^2\alpha \,  \dfrac{(1-z)(1-\tilde{z})}{1-z \, \tilde{z}}   + \cos^2\alpha &  .
\end{array}
\end{equation}

The four-form field strength of 11D supergravity has both a spacetime and an internal piece, namely,
\begin{equation}
\label{F4_all}
\hat{F}_{(4)} = \hat{F}^{\textrm{st}}_{(4)} + \hat{F}^{\textrm{tr}}_{(4)} \ .
\end{equation}
The spacetime part takes the form
\begin{equation}
\label{F4_st}
\hat{F}^{\textrm{st}}_{(4)} =  g \, h \, \textrm{vol}_{4} + g^{-1}  \sin (2\alpha) \,  \tilde{h}^{(3)} \wedge d\alpha  \ ,
\end{equation}
in terms of a function $\,h\,$ and a three-form $\,\tilde{h}^{(3)}\,$ given by
\begin{equation}
\begin{array}{rcl}
h 
&=&  i \, \dfrac{3 - z \, \tilde{z} + ( z + \tilde{z} ) \, \cos(2 \alpha)}{(1-z \tilde{z})} \ , \\[4mm]
\tilde{h}^{(3)}  &=&  i \,  \dfrac{(1-z^{2}) *_{4} d\tilde{z}+(1-\tilde{z}^{2}) *_{4}  dz}{(1- z \, \tilde{z})^{2}} \ .
\end{array}
\end{equation}
In the above expression, $*_{4}$ denotes the Hodge dual with respect to the external spacetime metric (\ref{ds4_ansatz}). The internal part is given by
\begin{equation}
\label{F4_tr}
\hat{F}^{\textrm{tr}}_{(4)} =  d\hat{A}^{\textrm{tr}}_{(3)}
\hspace{3mm} \textrm{ with } \hspace{3mm}
\hat{A}^{\textrm{tr}}_{(3)} = g^{-3} \, h_{1} \,  \textrm{vol}_{1} +  g^{-3} \, h_{2} \,  \textrm{vol}_{2} \ ,
\end{equation}
in terms of the functions
\begin{equation}
\label{h_functions_uplift}
h_{1}  =   i \, 8 \,  \dfrac{\cos^4\alpha}{f_{1}} \,  \dfrac{z-\tilde{z}}{1- z \tilde{z}}
\hspace{3mm} , \hspace{3mm}
h_{2}  =  i \, 8 \,  \dfrac{\sin^4\alpha}{f_{2}} \,  \dfrac{z-\tilde{z}}{1- z \tilde{z}}    \ ,
\end{equation}
and the volume forms on the two three-spheres $\textrm{vol}_{1,2}$. Note that $\hat{F}^{\textrm{tr}}_{(4)}$ is non-zero whenever the \textit{axionic} combination $z-\tilde{z}$ is non-zero. 

Due to the consistency of the truncation, any solution in the $\textrm{SO}(4) \times \textrm{SO}(4)$ invariant sector of the $\textrm{SO}(8)$ gauged supergravity is also a solution of the Euclidean equations of motion in (\ref{EOM_11D_sugra_Euclidean}) and satisfies the Bianchi identity (\ref{BI_11D_sugra}). Note also that, under the exchange $z \rightarrow \tilde{z}$, the 11D geometry in (\ref{ds_11D}) remains invariant whereas the four-form flux in (\ref{F4_all}) changes as $\hat{F}^{\textrm{st}}_{(4)} \rightarrow \hat{F}^{\textrm{st}}_{(4)}$ and $\hat{F}^{\textrm{tr}}_{(4)} \rightarrow -\hat{F}^{\textrm{tr}}_{(4)}$.

\subsection{$\textrm{SU}(3) \times \textrm{U}(1)^2$ invariant sector}
\label{sec:11D_SU(3)xU(1)xU(1)}

When the three pairs $(z_{i},\tilde{z}_{i})$ of scalar fields in the STU-model are identified, \textit{i.e.} $(z,\tilde{z}) \equiv (z_{i},\tilde{z}_{i})$ $\forall i$, the corresponding solutions belong to the $\textrm{SU}(3) \times \textrm{U}(1)^2$ invariant sector of the maximal $\textrm{SO}(8)$ supergravity. The uplift of this sector to eleven-dimensional supergravity has been worked out in \cite{Pilch:2015dwa} (see also \cite{Larios:2019kbw} whose conventions we closely follow) in the case of Lorentzian signature. As before, we extend the uplift formulae therein to Euclidean signature and adapt it to the unit-disk parameterisation of the scalar fields $z$ and $\tilde{z}$ we have adopted.

The eleven-dimensional (warped) geometry is of the form $\mathcal{M}_{4} \times \textrm{S}^{7}$ with $\textrm{S}^{7}=\mathcal{I} \times \mathbb{CP}_{2} \times \textrm{S}_{\tau}^{1} \times \textrm{S}_{\psi}^{1}$. The $\mathbb{CP}_{2} $ and $\textrm{S}_{\tau}^{1} \times \textrm{S}_{\psi}^{1}$ factors in the internal geometry are responsible for the $\textrm{SU}(3) \times \textrm{U}(1)^2$ residual symmetry of the four-dimensional solutions. Denoting $\alpha \in[0,\frac
{\pi}{2}]\,$ the coordinate along the interval $\mathcal{I}$, the 11D metric takes the form
\begin{equation}
\label{11D_metric_SU(3)xU(1)xU(1)}
\begin{array}{lll}
d\hat{s}_{11}^{2} & = & \frac{1}{2} \, f_{1} \, ds_{4}^{2} + 2 \,  g^{-2} \Big[ f_{2} \, d\alpha^{2} \\[2mm]
&+& \cos^{2}\alpha \, \big( \, f_{3} \, \, ds^{2}_{\mathbb{CP}_{2}} + \sin^{2}\alpha \, f_{4} \, (d\tau + \sigma)^{2} \, \big)\\[2mm]
& + & f_{5} \, \big( d\psi + \cos^{2} \alpha \, f_{6} \, (d\tau + \sigma) \big)^{2} \Big] \ ,
\end{array}
\end{equation}
with $\tau,\psi \in[0,2 \pi]$. We have denoted $\sigma$ the one-form on $\mathbb{CP}_{2}$ such that $d\sigma=2\boldsymbol{J}$ with $\boldsymbol{J}$ being the K\"ahler form on $\,\mathbb{CP}_{2}$. The metric in (\ref{11D_metric_SU(3)xU(1)xU(1)}) is fully specified in terms of six functions $f_{1\ldots 6}$ given by
\begin{equation}
\label{f_functions_SU(3)xU(1)xU(1)}
\begin{array}{c}
f_{1} = (1-z\tilde{z})^{-1} L^{\frac{1}{3}} \, H^{\frac{2}{3}}
\hspace{3mm} , \hspace{3mm}
f_{2} = f_{3}^{-2} = H^{\frac{2}{3}} L^{-\frac{2}{3}}  \ , \\[2mm]
f_{4} = (1-z\tilde{z})^{2}L^{-\frac{2}{3}} \, H^{\frac{2}{3}} \, K^{-1}   
\hspace{3mm} , \hspace{3mm}
f_{5} = H^{-\frac{4}{3}} \, K \, L^{-\frac{2}{3}} \ , \\[2mm]
f_{6} = \Big[ L \, H + ( z - \tilde{z})^{2} \cos(2\alpha) \Big] \, K^{-1} \ ,
\end{array}
\end{equation}
with
\begin{equation}
\begin{array}{ccll}
H &=& 1+z\tilde{z} - ( z +\tilde{z}) \cos(2\alpha) & , \\[2mm]
K &=& 1+(z\tilde{z})^{2} - 2 \, z\tilde{z} \, \cos(4\alpha) & , \\[2mm]
L &=& (1+z) (1+\tilde{z}) & .
\end{array}
\end{equation}

The four-form field strength of 11D supergravity consists of a spacetime and an internal piece
\begin{equation}
\hat{F}_{(4)} = \hat{F}^{\textrm{st}}_{(4)} + \hat{F}^{\textrm{tr}}_{(4)} \ .
\end{equation}
The spacetime part reads
\begin{equation}
\hat{F}_{(4)}^{\textrm{st}} =
g \, h_{1} \, \text{vol}_{4} + g^{-1} \, \sin(2 \alpha) \,\, h^{(3)}_{2} \wedge d\alpha \ ,
\end{equation}
in terms of a zero-form $h_{1}$ and a three-form $h_{2}^{(3)}$ given by
\begin{equation}
\begin{array}{rcll}
h_{1} &=& i \, \dfrac{3 \, (1+z\tilde{z}) + ( z +\tilde{z}) \, (1 - 2 \cos(2 \alpha) )}{2\sqrt{2} \, (1-z\tilde{z})} & , \\[4mm]
h_{2}^{(3)}  &=& i \, \dfrac{(1-z^{2}) *_{4}  d\tilde{z} + (1-\tilde{z}^{2}) *_{4} dz}{\sqrt{2}\,(1-z\tilde{z})^{2}}  & .
\end{array}
\end{equation}
Again, $*_{4}$ denotes the Hodge dual with respect to the external spacetime metric (\ref{ds4_ansatz}). Lastly, the internal part is given by
\begin{equation}
\begin{array}{lll}
\hat{F}_{(4)}^{\textrm{tr}} &=& 2 \sqrt{2} \,  g^{-3} \Big[  \, \sin(2 \alpha) \, h^{(1)}_{3} \wedge d\alpha \wedge d\psi \wedge(d\tau+\sigma)\\[2mm]
& + &   \cos^{4} \alpha \,\, h^{(1)}_{4} \wedge (d\tau + \sigma) \wedge\boldsymbol{J} \\[2mm]
& + & \cos^{2} \alpha \, \cos(2 \alpha) \,\, h_{5}^{(1)} \wedge d\psi \wedge\boldsymbol{J}\\[2mm]
& + &   \cos^{2}\alpha \sin(2 \alpha) \, h_{6} \, d\alpha \wedge d\psi \wedge\boldsymbol{J} +   \cos^{4} \alpha \,\, h_{7} \, \boldsymbol{J} \wedge\boldsymbol{J}\\[2mm]
& + &   \cos^{2}\alpha \, \sin(2 \alpha) \,\, h_{8} \, d\alpha \wedge(d\tau + \sigma) \wedge\boldsymbol{J} \, \Big] \ ,
\end{array}
\end{equation}
in terms of the one-forms
\begin{equation}
\label{11D_F4_one-forms}
\begin{array}{lll}
h^{(1)}_{3} & = & \dfrac{i}{2} \left(  \dfrac{d\tilde{z} }{(1+\tilde{z})^{2}} - \dfrac{dz }{(1+z)^{2}} \right)  \ ,\\[4mm]
h^{(1)}_{4} & = & h^{(1)}_{5} \,\, =  \, \dfrac{i}{2H^{2}} \, \Big[ \,\,\left(  1 - 2 \cos(2 \alpha) \, \tilde{z} + \tilde{z}^{2} \right)  \, dz \\
& & \hspace{20mm} - \left(  1 - 2 \cos(2 \alpha) \, z + z^{2} \right)  \, d\tilde{z}
\,\, \Big] \ ,
\end{array}
\end{equation}
and the zero-forms
\begin{equation}
\begin{array}{lll}
h_{6} & = & i \, 2 \, (1+z\tilde{z}) \, \dfrac{(\tilde{z}-z)}{L H^2} \, \Big( 1+ z\tilde{z} + (z+\tilde{z}) \sin^{2}\alpha \Big) \ ,\\[4mm]
h_{7} & = & - i \, \dfrac{(\tilde{z}-z)}{2 H} \ ,\\[4mm]
h_{8} & = & i \,\dfrac{(\tilde{z}-z)}{H^2} \, \Big( 1+ z\tilde{z} + (z+\tilde{z}) \sin^{2}\alpha \Big) \ .
\end{array}
\end{equation}
Note that the purely internal part of $\hat{F}^{\textrm{tr}}_{(4)}$ is non-zero whenever the \textit{axionic} combination $z-\tilde{z}$ is non-zero.

By virtue of the consistency of the truncation, any solution in the $\textrm{SU}(3) \times \textrm{U}(1)^2$ invariant sector of the $\textrm{SO}(8)$ gauged supergravity is also a solution of the Euclidean equations of motion in (\ref{EOM_11D_sugra_Euclidean}) and satisfies the Bianchi identity (\ref{BI_11D_sugra}).

\bibliographystyle{utphys}
\bibliography{Biblio}

\end{document}